\documentclass[a4paper,11pt]{article}
\usepackage{pos}
\usepackage{float}
\usepackage[symbol]{footmisc}
\renewcommand{\thefootnote}{\fnsymbol{footnote}}
\newcommand{\KK}{${\cal KK}$}
\def\rQCED{{\rm QCED}}
\newcommand{\qb}{{\bar{q}}}
\newcommand{\sfac}{\mathfrak{s}}
\title{Amplitude-Based IR-Improvement in Precision LHC$/$FCC Physics}
\centerline{BU-HEPP-24-03, 09/2024}
\author*[a]{B.F.L. Ward}
\author[b]{S. Jadach\footnote[2]{Deceased.}} 
\author[c]{W. Placzek}
\author[b]{M. Skrzypek}
\author[b]{Z. A. Was}
\author[d]{S.A. Yost}

\affiliation[a]{Department of Physics, Baylor University,\\
  Waco, TX, USA}

\affiliation[b]{Institute of Nuclear Physics,\\
Krakow, PL}

\affiliation[c]{Institute of Applied Computer Science, Jagiellonian University,\\
Krakow, PL}

\affiliation[d]{Department of Physics, The Citadel,\\
  Charleston, SC, USA}

\emailAdd{bfl\_ward@baylor.edu}
\emailAdd{wieslaw.placzek@uj.edu.pl}
\emailAdd{maciej.skrzypek@ifj.edu.pl}
\emailAdd{zbigniew.was@ifj.edu.pl}
\emailAdd{yosts1@citadel.edu}

\abstract{We present recent results based on the IR-improvement of unintegrable singularities in the infrared regime via amplitude-based resummation in $QED \times QCD \subset SU(2)_L \times U_1 \times SU(3)_c$. In the context of precision LHC/FCC physics, we focus on specific examples, such as the removal of QED contamination in PDF’s evolved from data at $Q_0^2\sim 2 GeV^2$  and used in the evaluating precision observables in $pp\rightarrow Z + X \rightarrow \ell\bar\ell +  X’$, in which we discuss new results and new issues.}

\FullConference{42nd International Conference on High Energy Physics (ICHEP2024)\\
18-24 July 2024\\
Prague, Czech Republic\\}

\begin{document}
\maketitle

\baselineskip=10pt
\section{In Memory of Prof. Stanislaw Jadach}
\begin{figure}[h]
\begin{center}
\setlength{\unitlength}{1in}
\begin{picture}(6,4.5)(0,0)
\put(0.5,0){\includegraphics[width=5in]{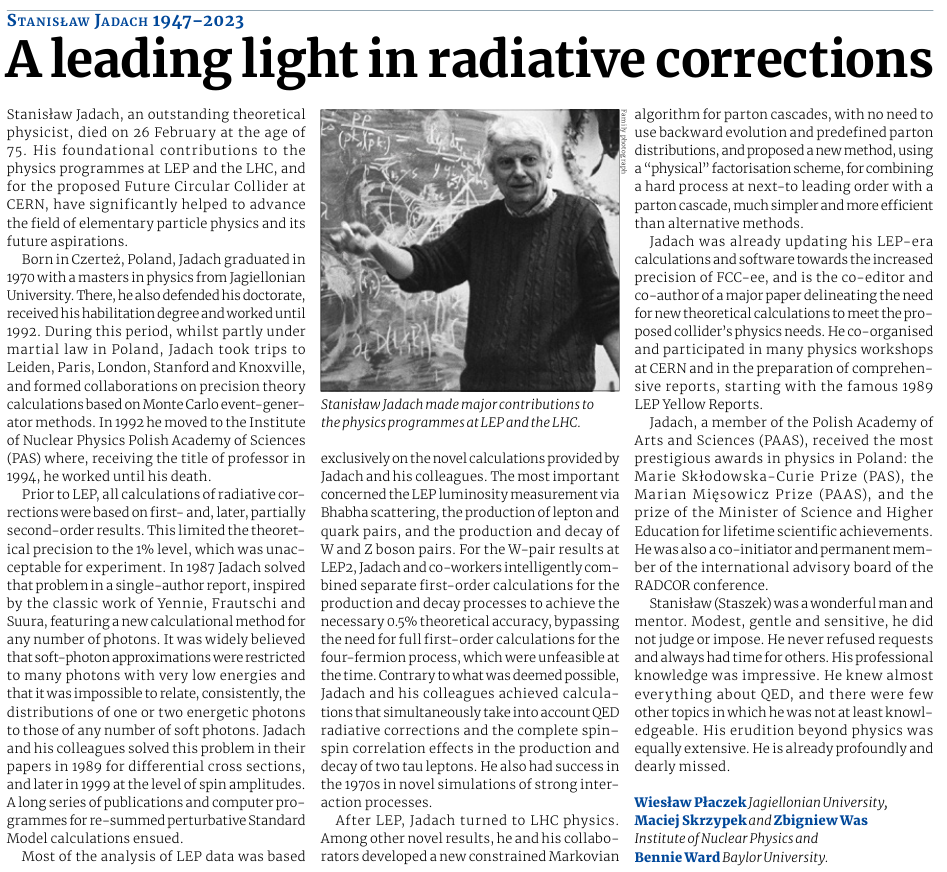}}
\end{picture}
\end{center}
\vspace{-5mm}
\caption{{\it CERN Courier} obituary for the late Prof. Stanislaw Jadach.}
\label{fig1}
\end{figure}

Sadly, on February 26, 2023 my (BFLW) close friend and collaborator, Prof. Stanislaw Jadach, passed away suddenly.Here in Fig.~\ref{fig1} we reproduce his {\it CERN Courier}\footnote{{\it CERN Courier}, May/June 2023 issue, p.59.} obituary. 
He was a pioneering researcher in the theory and MC simulation of higher order corrections in quantum field theory. As this is the first Rochester Conference since his passing, we lift up his special contributions to our field which helped to keep our field alive. We all miss him dearly. This contribution is dedicated in memoriam to him.\par

\section{Recapitulation of YFS Exact Amplitude-Based Resummation}
It is still true that our YFS~\cite{yfs:1961} exact amplitude-based resummation theory, especially its CEEX~\cite{Jadach:2000ir} realization, is generally familiar in our field. We present in this section a brief recapitulation of the theory accordingly before turning to some new results based on it in the next section.\par
The theory is exhibited by the following master formula:
{\small
\begin{eqnarray}
&d\bar\sigma_{\rm res} = e^{\rm SUM_{IR}(QCED)}
   \sum_{{n,m}=0}^\infty\frac{1}{n!m!}\int\prod_{j_1=1}^n\frac{d^3k_{j_1}}{k_{j_1}} \cr
&\prod_{j_2=1}^m\frac{d^3{k'}_{j_2}}{{k'}_{j_2}}
\int\frac{d^4y}{(2\pi)^4}e^{iy\cdot(p_1+q_1-p_2-q_2-\sum k_{j_1}-\sum {k'}_{j_2})+
D_\rQCED} \cr
&{\tilde{\bar\beta}_{n,m}(k_1,\ldots,k_n;k'_1,\ldots,k'_m)}\frac{d^3p_2}{p_2^{\,0}}\frac{d^3q_2}{q_2^{\,0}},
\label{subp15b}
\end{eqnarray}}
where the {\em new}\footnote{The {\em non-Abelian} nature of QCD requires a new treatment of the corresponding part of the IR limit~\cite{Gatheral:1983} so that we usually include in ${\rm SUM_{IR}(QCED)}$ only the leading term from the QCD exponent in Ref.~\cite{Gatheral:1983} -- the remainder is included in the residuals $\tilde{\bar\beta}_{n,m}$ .}(YFS-style) residuals   
{$\tilde{\bar\beta}_{n,m}(k_1,\ldots,k_n;k'_1,\ldots,k'_m)$} have {$n$} hard gluons and {$m$} hard photons. The new residuals and the  infrared functions ${\rm SUM_{IR}(QCED)}$ and ${ D_\rQCED}$ are defined in Ref.~\cite{mcnlo-hwiri,mcnlo-hwiri1}.  As explained in Ref.~\cite{mcnlo-hwiri,mcnlo-hwiri1}, parton shower/ME matching engenders the replacements {$\tilde{\bar\beta}_{n,m}\rightarrow \hat{\tilde{\bar\beta}}_{n,m}$}, which allow us to connect with  MC@NLO~\cite{mcnlo,mcnlo1}, via the basic formula{\small
\begin{equation}
{d\sigma} =\sum_{i,j}\int dx_1dx_2{F_i(x_1)F_j(x_2)} d\hat\sigma_{\rm res}(x_1x_2s).
\label{bscfrla}
\end{equation}}
\par
 Eq.(\ref{subp15b}) has been used to obtain new results in precision LHC and FCC physics. One of us (BFLW) has extended (See Ref.~\cite{ijmpa2018} and references therein.)  eq.(\ref{subp15b}) to general relativity as an approach to quantum gravity.  New results are accompanied with new perspectives in each of our applications, as we illustrate in the next Section.\par

\section{New Perspectives for Precision Collider Physics: LHC, FCC, CEPC, CPPC, ILC, CLIC}
\begin{figure}[h!]
\begin{center}
\setlength{\unitlength}{1in}
\begin{picture}(6,2.0)(0,0)
\put(0.5,0){\includegraphics[width=5in]{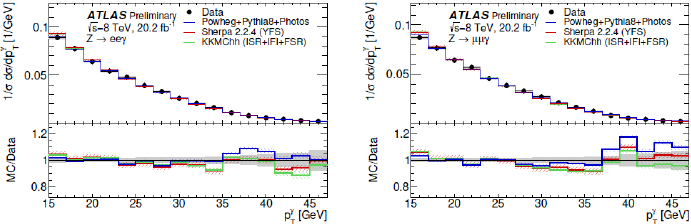}}
\end{picture}
\end{center}
\vspace{-5mm}
\caption{ATLAS analysis of $Z/\gamma$ production at $8$ TeV.}
\label{fig3}
\end{figure}

We have a new perspective on the expectations for precision physics for the Standard Theory
EW interactions at HL-LHC due to the realization of eq.(\ref{subp15b}) in the MC event generator \KK{MC}-hh~\cite{kkmchh1-sh} by four of us (SJ, BFLW, ZAW, SAY). We illustrate this here with the plots in Fig.~\ref{fig3} in the ATLAS analysis~\cite{atlas-epjc-zg:2024} of $Z\gamma$ production at 8 TeV.
The Powheg-Pythia8-Photos~\cite{powheg-org,powheg-orga,powheg1,powhega,Sjostrand:2007gs-sh,Golonka:2006tw}, Sherpa2.2.4(YFS)~\cite{sherpa,sherpa-2.2}, and \KK{MC}-hh predictions for the $\gamma p_T$ spectrum are compared with the ATLAS data. At this point, with the level of the uncertainties in the data, 
all three predictions are in reasonable agreement with the data. With ~100 times the featured statistics, at HL-LHC a precision test against the theories will obtain.\par
\begin{figure}[h!]
\begin{center}
\setlength{\unitlength}{1in}
\begin{picture}(6,2.0)(0,0)
\put(0.5,0){\includegraphics[width=5in]{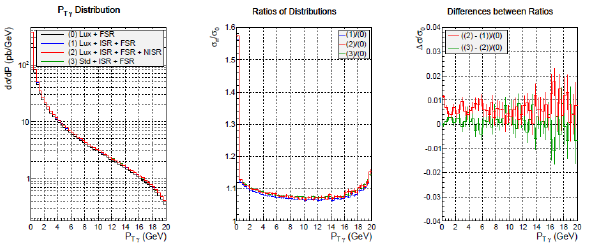}}
\end{picture}
\end{center}
\vspace{-5mm}
\caption{For events with at least one photon, the distribution for $P_{T_\gamma}$ of the photon for which it is greatest for events with each lepton having $ p_{T\ell}> 25 GeV, \eta_\ell< 2.5$ calculated with (0) FSR only (black). (1) FSR + ISR (blue). and (2) FSR + ISR with NISR (red) for NNPDF3.1-LuxQED NLO PDFs. For comparison, (3) shows FSR + ISR with ordinary NNPDF3.1 NLO PDFs (green). The center graph shows
ISR on/off ratios (1)/(0) (blue),(2)/(0) (red) and (3)/(0) (green). The right-hand graph shows the fractional differences ((1)- (2))/(0) in red and ((2)- (3))/(0) in green.}
\label{fig4}
\end{figure}

In a new perspective toward the important issue of the effect of QED contamination in non-QED PDFs ~\cite{jad-yost-afb,ichep2022-say,sjetaltoappear} we use Negative ISR (NISR) evolution to address the size of this contamination directly. Using a standard notation for PDFs and cross sections, we have the cross section representation
\begin{equation}
\begin{split}
\sigma(s)&=
\frac{3}{4}\pi\sigma_0(s)\!\!\!
\sum_{q=u,d,s,c,b} \int d\hat{x}\;  dz dr dt \; \int dx_q dx_\qb\; 
\delta(\hat{x}-x_qx_\qb z t)
\\&\times
f^{h_1}_q(   s\hat{x}, x_q) 
f^{h_2}_\qb( s\hat{x}, x_\qb) \;
 \rho_I^{(0)}\big(\gamma_{Iq}(s\hat{x}/m_q^2),z\big)\; 
 \rho_I^{(2)}\big(-\gamma_{Iq}(Q_0^2/m_q^2),t\big)\; 
\\&\times
\sigma^{Born}_{q\qb}(s\hat{x}z)\;
\langle W_{MC} \rangle,
\label{eq:kkhhsigmaPru}
\end{split}
\end{equation}
which includes an extra convolution with the well known second order exponentiated ISR
``radiator function'' $\rho_I^{(2)} $ with the negative evolution time argument
$-\gamma_{Iq}(Q_0^2/m_q^2)$ defined in Ref.~\cite{jad-yost-afb}. This removes the  QED below $Q_0$. This is exhibited in Fig.~\ref{fig4} from Ref.~\cite{ichep2022-say}  
for the $P_{T_\gamma}$ for the photon for which it is the largest in $Z\gamma^*$ production and decay to lepton pairs at the LHC at $8 ~TeV$ for
cuts as described in the figure. In agreement with arguments in Ref.~\cite{kkmchh2}, the results in the figure show that the effect of QED contamination in non-QED
PDFs is below the errors on the PDFs.\par
\begin{figure}[pt!]
\begin{center}
\setlength{\unitlength}{1in}
\begin{picture}(6.5,3.0)(0,0)
\put(0,1.5){\includegraphics[width=3.0in,height=1.5in]{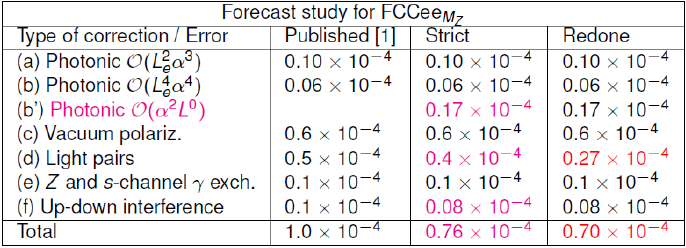}}
\put(3.05,1.5){\includegraphics[width=3.0in,height=1.5in]{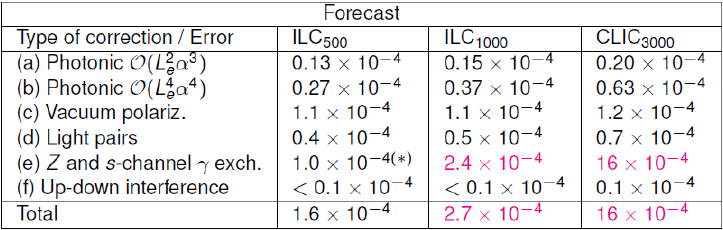}}
\put(1.5,1.3){(a)\hspace{2.8in}(b)}
\end{picture}
\end{center}
\vspace{-40mm}
\caption{\baselineskip=11pt Current purview on luminosity theory errors: (a), FCC-ee at $M_Z$; (b), proposed higher energy colliders}
\label{fig5}
\end{figure}

In view of the planned EW/Higgs factories, five of us (SJ, WP, MS, BFLW, SAY) have discussed in Refs.~\cite{Jadach:2018jjo,Jadach:2021ayv-sh,fcc2023wkshpms,skrzypek-2023mttd} the new perspectives for the BHLUMI~\cite{bhlumi4:1996} luminosity theory error. In Fig.~\ref{fig5}~\cite{fcc2023wkshpms}, to illustrate this new perspective,  we show the current purview for the FCC-ee at $M_Z$ and for the proposed higher energy colliders. 
We note the improvements at $M_Z$ shown in Fig.~\ref{fig5}(a) to 0.007\% and that the use of the results in Ref.~\cite{fjeger-fccwksp2019} together with lattice methods~\cite{latt1,latt2,latt3}
opens the possibility that item (c) in Fig~\ref{fig5}(a) could be reduced by a factor of ~ 6\footnote{The formula to be studied is $\Delta\alpha_{had}(t)=\Delta\alpha_{had}(-Q^2_0)|_{lat}+[\Delta\alpha_{had}(t)-\Delta\alpha_{had}(-Q^2_0)]|_{pQCDAdler}$ with {\it lat} denoting the methods of Refs.~\cite{latt1,latt2,latt3} and {\it pQCDAdler} denoting the methods of Ref.~\cite{fjeger-fccwksp2019}.}. \par
\section{Improving the Collinear Limit in YFS Theory}
The higher precision requirements for HL-LHC/FCC physics motivate improving the collinear limit of YFS theory. For,
it is known~\cite{Jadach:2023cl1} that, for the process $e^+(p_2)\;e^-(p_1)\rightarrow \bar{f}(p_4)\;e^-(p_3)$, in the usual YFS theory the s-channel virtual infrared function $B$ resums (exponentiates) the non-infrared term
$\frac{1}{2}Q_e^2{\alpha\over\pi} L$ whereas the methods in Ref.~\cite{gribv-lptv:1972} show that the term $\frac{3}{2}Q_e^2{\alpha\over\pi} L$ exponentiates -- see also Refs.~\cite{frixione-2019,bertone-2019,frixione-2021,bertone-2022} for recent developments in the attendant collinear factorization approach. Here, we use an obvious notation where the respective big log is $L = \ln(s/m_e^2)$ when $s=(p_1+p_2)^2$ is the center-of-mass energy squared. In Ref.~\cite{Jadach:2023cl1}, three of us (SJ, BFLW, ZAW) have shown that the suggested collinear enhancement of YFS theory which exponentiates the term $\frac{3}{2}Q_e^2{\alpha\over\pi} L$ does exist.\par
Specifically, we find that the virtual infrared function $B$ in the s-channel and the corresponding real infrared function $\tilde{B}$ can be extended to {\small
\begin{equation}
\begin{split}
B_{CL}&\equiv B+{\bf \Delta B}\\
          &= \int {d^4k\over k^2} {i\over (2\pi)^3} 
                       \bigg[\bigg( {2p-k \over 2kp-k^2} - {2q+k \over 2kq+k^2} \bigg)^2{\bf -\frac{4pk-4qk}{(2pk-k^2)(2qk+k^2)}}\bigg],
\end{split}
\end{equation}}
 {\small
\begin{equation}
\begin{split}
\tilde{B}_{CL} &\equiv \tilde{B}+{\bf \Delta\tilde{B}}\\
 &=\frac{-1}{8\pi^2}\int\frac{d^3k}{k_0}\bigg\{\large(\frac{p_1}{kp_1} - \frac{p_2}{kp_2}\large)^2 +{\bf \frac{1}{kp_1}\large(2 -\frac{kp_2}{p_1p_2}\large)}
                                          +{\bf \frac{1}{kp_2}\large(2 -\frac{kp_1}{p_1p_2}\large)}\bigg\},
\end{split}
\label{eq-real2}
\end{equation}}
\noindent where the extensions are indicated in boldface in an obvious notation. These extensions leave the YFS infrared algebra unaffected while the $B_{CL}$ does exponentiate the entire $\frac{3}{2}Q_e^2{\alpha\over\pi} L$ term and the $\tilde{B}_{CL}$ does carry the respective collinear big log of the exact result in Ref.~\cite{berends-neerver-burgers:1988} in the soft regime.\par 
In the CEEX case, the corresponding collinear extension of the soft eikonal amplitude factor defined in Ref.~\cite{ceex2:1999sh}
for the photon polarization $\sigma$ and $e^-$  helicity $\sigma'$ is given by {\small
\begin{equation}
\begin{split}
\sfac_{CL,\sigma}(k) = \sqrt{2}Q_ee\bigg[-\sqrt{\frac{p_1\zeta}{k\zeta}}\frac{<k\sigma|\hat{p}_1 -\sigma>}{2p_1k}
    +{\bf \delta_{\sigma'\;-\sigma}\sqrt{\frac{k\zeta}{p_1\zeta}}\frac{<k\sigma|\hat{p}_1 \sigma'>}{2p_1k}}\\
    + \sqrt{\frac{p_2\zeta}{k\zeta}}\frac{<k\sigma|\hat{p}_2 -\sigma>}{2p_2k}+{\bf \delta_{\sigma' \sigma}\sqrt{\frac{k\zeta}{p_2\zeta}}\frac{<\hat{p}_2 \sigma'|k -\sigma>}{2p_2k}}\bigg],
\end{split}
\label{eq-real4}
\end{equation}}
where from Ref.~\cite{ceex2:1999sh} $\zeta\equiv (1,1,0,0)$ for our choice for the respective auxiliary vector in our Global Positioning of Spin (GPS)~\cite{gps:1998} spinor conventions with the consequent definition 
$\hat{p}= p - \zeta m^2/(2\zeta p)$
for any four vector $p$ with $p^2 = m^2.$ The collinear extension terms are again indicated in boldface.\par
These extended infrared functions are expected to give in general a higher precision for a given level of exactness~\cite{bflwetaltoappear}.\par
\baselineskip=10pt
\bibliography{Tauola_interface_design}{}
\bibliographystyle{utphys_spires}

\end{document}